# Probing the anomalous symmetry-breaking in kagome material $CsV_3Sb_5$ via third-order nonlinearity

Zheng Dai[1,#], Fengyi Guo[1,#], Shuai Zhang[1,2,3,*], Xiubing Li[1], Congcong Li[1], Yufan Ju[1], Ziqi Wang[1], Bin Cheng[1,4], Zhe Ying[3], Fengqi Song[1,2,3,*]

[1] National Laboratory of Solid-State Microstructures, Collaborative Innovation Center of Advanced Microstructures, Jiangsu Physical Science Research Center, and School of Physics, Nanjing University, Nanjing 210093, China

[2] Institute of Atom Manufacturing, Nanjing University, Nanjing 211800, China

[3] Jiangsu Provincial Key Laboratory of Atomic Level Manufacturing, Nanjing Institute of Atomic Scale Manufacturing, Nanjing 211800, China

[4] School of Physics and Electronic Information, Jiangsu Second Normal University, Nanjing 211200, China

#These authors contributed equally to this work.

*Corresponding authors. S.Z. (szhang@nju.edu.cn), F.S. (songfengqi@nju.edu.cn)

**Abstract**

The kagome material has rapidly established itself as a research frontier in condensed matter physics, owing to its distinctive geometric structure and the rich array of unconventional physical phenomena. In the kagome $AV_3Sb_5$ (A = K, Rb, Cs) family, the charge-ordered state exhibits a remarkable characteristic, i.e., anomalous symmetry-breaking, which is tied to the topological nature of the electronic band structure. Here, we report the third-order nonlinear longitudinal and Hall responses that persist stably up to room temperature in the kagome material $CsV_3Sb_5$. Notably, the nonlinear responses demonstrate significant enhancement below the charge density wave (~ 77 K) order and anomalous symmetry-breaking (~ 39 K) state. The scaling analysis indicates that the third-order nonlinear transport is governed jointly by quantum geometric contribution and extrinsic scattering. This study realizes a giant third-order nonlinear response and provides a distinct method to detect anomalous symmetry-breaking in $CsV_3Sb_5$.

## Introduction

Kagome lattice materials, characterized by their distinctive geometric configurations and atomic arrangements, exhibit unique electronic structures[1-5] that include flat bands, Dirac points, and saddle points. These properties render kagome lattice materials an ideal platform for the exploration of topological physics and correlated electron physics[6-8]. The recently discovered quasi-two-dimensional kagome $AV_3Sb_5$ (A = K, Rb, Cs)[9] possesses a Fermi level located near the electronic saddle point. This peculiar band structure enables the coexistence of nontrivial topological electronic states[10-12], charge density wave (CDW) order[13], and superconductivity[13], which is rare among conventional kagome systems. Consequently, the $AV_3Sb_5$ system provides an exceptional arena for investigating emergent electronic states and the complex interplay, garnering widespread attention[14].

As a member of the $AV_3Sb_5$ family, bulk $CsV_3Sb_5$ exhibits a superconducting transition temperature ($T_C$) of 2.5 K and forms a 2×2×2 CDW order below 94 K[13]. A prominent competition exists between the superconductivity and CDW order of $CsV_3Sb_5$[15-19], and both $T_C$ and CDW transition temperature ($T_{CDW}$) can be effectively tuned by the sample thickness[18, 19] and external pressure[17]. In $CsV_3Sb_5$, the formation of CDW order is accompanied by the simultaneous emergence of electronic nematicity[20-23] and magnetic-field-induced nonlinearity of Hall effect[12, 24-26]. As the system cools toward the superconducting regime, exotic superconducting phenomena emerge[27-31]. Notably, in the intermediate temperature range between $T_C$ and $T_{CDW}$, $CsV_3Sb_5$ hosts abundant anomalous phenomena[32-37], such as electronic nematicity[38], anomalous Nernst effect[39], $4a_0$ charge order[40], twofold resistivity anisotropy[20, 21], time-reversal symmetry breaking[31, 34, 41, 42], rotational symmetry breaking[23, 43, 44], and electronic magnetochiral anisotropy (eMCA) transport[45]. Despite extensive research[40, 46-48], particularly around 35 K, many fundamental questions concerning the anomalous symmetry breaking in CsV□Sb□ remain open.

As one of the research hotspots in condensed matter physics, the nonlinear transport is highly sensitive to the symmetry of materials[49, 50]. Theoretical and experimental studies on the nonlinear transport have confirmed that it is closely related to the non-centrosymmetric structure[51-53], topological properties[54, 55], chirality[54], CDW[56, 57], and antiferromagnetic phase transition[58-60]. Therefore, the nonlinear transport is expected to be an effective tool for probing the complex anomalous symmetry-breaking in $CsV_3Sb_5$.

In this work, we report a giant third-order nonlinear responses in kagome material $CsV_3Sb_5$ that persist up to room temperature, while the second-order nonlinear response is negligible. Intriguingly, we observe the coexistence of longitudinal and Hall third-order nonlinear responses, with the longitudinal component exhibiting a magnitude more than four times that of the Hall counterpart. Furthermore, the third-order nonlinear responses exhibit two distinct enhancements with decreasing temperature: a prominent enhancement emerges at $T_{CDW}$ ~ 77 K, corresponding to the CDW phase transition, followed by a noticeable increase below $T^*$ ~ 39 K, originating from the anomalous symmetry-breaking. Combined with the scaling analysis, we conclude that below $T_{CDW}$, the third-order nonlinear longitudinal transport is governed by the joint contribution of the quantum metric quadrupole and extrinsic effects, and the third-order nonlinear Hall transport arises from the Berry curvature quadrupole together with extrinsic effects. By contrast, both the third-order nonlinear responses are dominated by extrinsic effects above $T_{CDW}$.

## Results and Discussions

The $CsV_3Sb_5$ flake device is fabricated via electron beam lithography and electron beam evaporation, as shown in the inset of Fig. 1a. The thickness of the $CsV_3Sb_5$ flake measured by atomic force microscopy is 22.7 nm (Supporting Information Fig. S1). The temperature-dependent longitudinal resistance of the $CsV_3Sb_5$ device is presented in Fig. 1a. The $CsV_3Sb_5$ enters the superconducting state at 4.5 K, while a CDW transition occurs at 77 K. Compared with bulk $CsV_3Sb_5$[13], the thin flake device exhibits a slightly higher $T_C$ and a lower $T_{CDW}$. According to previous studies, when the $CsV_3Sb_5$ flake is thinned from bulk to the atomic limit, $T_C$ first increases and then decreases. In contrast, $T_{CDW}$ follows an opposite evolution: it decreases first and then rises. This behavior originates from a crossover from the weakening electron-phonon coupling to the enhanced electronic interactions as the thickness of $CsV_3Sb_5$ decreases[19] (see Supporting Information Fig. S13 for detailed discussions of the higher $T_C$).

We then study the magnetic transport properties of $CsV_3Sb_5$ flake. The magnetic field dependence of longitudinal resistance $R_{\parallel}$ and step-like Hall resistance $R_{\perp}^{SL}$ under different temperatures are shown in Figs. 1b and 1c, respectively. The longitudinal resistance and Hall resistance are symmetrized and antisymmetrized, respectively, to eliminate the mixing effect. The step-like Hall resistance $R_{\perp}^{SL}$ is obtained from the Hall resistance $R_{\perp}$ (Supporting Information Fig. S10) by subtracting the linear background $R_{\mathrm{n}}$. The step-like Hall resistance of $CsV_3Sb_5$ persists up to $T_{CDW}$ and vanishes above $T_{CDW}$, demonstrating a close relation between CDW and magnetic-field-induced nonlinearity of Hall effect[12, 24-26]. Diverse physical mechanisms have been put forward to explain the observed nonlinearity of Hall effect. One viewpoint holds that the it is anomalous Hall effect (AHE) [12, 16], which is caused by the enhanced skew scattering in the CDW state and the large Berry curvature. By contrast, an alternative interpretation holds that this behavior represents an ordinary Hall anomaly solely arising from the tiny Fermi pockets with high mobility carriers[26].

In kagome material $CsV_3Sb_5$, the breaking of inversion symmetry and time-reversal symmetry driven by the CDW order gives rise to abundant exotic properties[20-23]. In particular, the anomalous symmetry-breaking at ~35 K has drawn extensive attention in the transport studies of $CsV_3Sb_5$[21, 39, 45, 47]. Meanwhile, the nonlinear transport, which has been widely investigated in recent years, has been proven to be a powerful probe for detecting symmetry characteristics in quantum materials[50, 51]. Therefore, we investigate the anomalous symmetry-breaking in $CsV_3Sb_5$ via the nonlinear transport. Figure 2a shows the linear dependence of the first-order longitudinal voltage $V_{\parallel}$ and Hall voltage $V_{\perp}$ on the amplitude of the alternating current $I^{\omega}$ at 5 K and zero magnetic field. The linearity of the first-order voltages confirms the good ohmic contacts of the $CsV_3Sb_5$ device, and the corresponding linear DC voltage-current characteristics are presented in Supporting Information Fig. S7.

Figure 2b depicts the current dependence of the second- and third- order longitudinal and Hall voltages. Both the third-order nonlinear longitudinal voltage $V_{\parallel}^{3\omega}$ and Hall voltage $V_{\perp}^{3\omega}$ exhibit a cubic dependence on the current, while the second-order longitudinal $V_{\parallel}^{2\omega}$ and Hall $V_{\perp}^{2\omega}$ voltages are negligible, demonstrating that the third-order nonlinearity is dominant. And the amplitude of the third-order nonlinear longitudinal voltage is more than four times that of the third-order nonlinear Hall voltage. Meanwhile, in Figs. 2c and 2d, the third-order nonlinear longitudinal and Hall signals remain unchanged when both the current and voltage leads are reversed, which are typical signatures of third-order nonlinearity[49, 61]. Furthermore, the frequency-dependent third-order nonlinear transport measurements, as shown in Figs. 2e and 2f, demonstrate that both the third-order nonlinear longitudinal and Hall voltages are independent of the frequencies, which rules out the influence of the spurious capacitive coupling effect on the third-order nonlinear signals. Additionally, we have included detailed discussions to rule out the potential influences of contact junction effects and thermoelectric effects, further verifying the reliability of the observed third-order nonlinear signals (Supporting Information Figs. S2-S8).

To analyze the third-order nonlinearity in $CsV_3Sb_5$ and investigate the relationship between anomalous symmetry-breaking and the third-order nonlinear transport in $CsV_3Sb_5$, the temperature-dependent third-order nonlinear transport have been measured, as shown in Figs. 3a and 3b. The third-order nonlinear signals remain prominent even at 300 K. Across the entire temperature range, both the third-order nonlinear longitudinal voltages and Hall voltages show linear relationship with the cube of the first-order longitudinal voltages, and the first-order longitudinal voltages maintain a good linear relationship from 5 K to 300 K (Supporting Information Fig. S9).

The third-order nonlinear longitudinal and Hall voltages at each temperature are shown in Figs. 3c and 3d, respectively. The amplitudes of both the third-order nonlinear longitudinal and Hall signals initially increase slowly as the temperature decreases. Subsequently, the nonlinear signals increase rapidly and reach the maximum at 77 K, after which they decrease again and exhibit the minimum at 39 K. Upon further cooling, they exhibit another prominent increase. Interestingly, there are two distinct kinks at 77 K and 39 K. The kink at 77 K can be readily assigned to the CDW transition of $CsV_3Sb_5$. Similar anomalies in third-order nonlinear signals near $T_{CDW}$ have also been reported in 1*T*-phase $VSe_2$[57]. The kink near 39 K is related to the anomalous symmetry-breaking in $CsV_3Sb_5$, which can be attributed to the electronic nematic transition[38]. As mentioned above, several experimental evidences have confirmed the occurrence of anomalous symmetry-breaking in $CsV_3Sb_5$ at ~35 K[20, 21, 33, 38, 39, 45, 62]. We have summarized our experimental results together with previous reports in Supporting Information Table S1. It can be seen that the characteristic temperature of the anomalous symmetry-breaking observed here is consistent with previous reports by other experimental techniques[33, 36, 38, 39, 45, 47, 62], while the $CsV_3Sb_5$ flake exhibits a reduced CDW transition temperature compared with bulk counterparts. This demonstrates that the third-order nonlinearity can serve as an effective probe for the anomalous symmetry-breaking in $CsV_3Sb_5$ (see Supporting Information Table S1 for detailed discussions). Moreover, the kink in the temperature dependent third-order nonlinearity enables a relatively precise determination of $T^*$ and $T_{CDW}$.

Below $T_{CDW}$, the unique symmetry characteristic of $CsV_3Sb_5$ demonstrates that the third-order nonlinear transport behaviors are closely associated with intrinsic quantum geometric effects, including the quantum metric and Berry curvature. As presented in Supporting Information Fig. S10, the third-order nonlinear longitudinal signal exhibits an even dependence on the magnetic field, whereas the third-order nonlinear Hall signal shows an odd magnetic-field dependence. It is consistent with previous theoretical and experimental results[49, 58, 59, 63, 64], in which the quantum metric quadrupole induces third-order longitudinal nonlinearity and the Berry curvature quadrupole dominates third-order Hall nonlinearity.

Given the crucial role of scattering in nonlinear transport, the scaling analyses of the third-order nonlinear transport are performed to distinguish the intrinsic quantum geometric contribution and extrinsic contribution. Following the analytical methods for conventional AHE[65] and second-order nonlinear Hall effect (NLHE)[66], the scaling behavior of the third-order nonlinear response is characterized by the variation of the ratio $V^{3\omega}/V_{\parallel}^{3}$ with the longitudinal conductivity $\sigma$. Figures 4a and 4b exhibit the temperature evolution of the $\sigma$ and the $V^{3\omega}/V_{\parallel}^{3}$. The third-order nonlinear voltage ratios decrease monotonically. The magnitude at low temperatures varies by approximately 3000 times compared with that at room temperature (Fig. 4b). A classical scaling formula for the third-order nonlinear longitudinal and Hall transport is adopted for fitting[58, 59, 67, 68]: $V^{3\omega}/\sigma V_{\parallel}^{3} = \alpha\sigma^2 + \beta$, where $\alpha$ represents the extrinsic skew scattering contribution and $\beta$ corresponds to the intrinsic quantum geometric contribution. The experimental data can be well fitted by the formula and the corresponding scaling fitting results are presented in Figs. 4c and 4d. The third-order nonlinear transport behavior is distinctly divided into three regimes bounded at 39 K and 77 K, which are consistent with the results shown in Figs. 3c and 3d. Notably, above $T_{CDW}$, the third-order nonlinear signals nearly disappear as the $\sigma$ approaches zero, corresponding to a near-zero value of $\beta$. This indicates that the third-order nonlinear responses are entirely dominated by extrinsic skew scattering above $T_{CDW}$. Below $T_{CDW}$, the intrinsic quantum geometric contribution to the third-order nonlinear transport is proportional to the $\sigma$, while the extrinsic skew scattering contribution scales with the cube of $\sigma$. The detailed results of the intrinsic and extrinsic contribution are presented in Supporting Information Fig. S11. We find that the extrinsic skew scattering plays an indispensable role in the generation of the third-order nonlinear transport across the entire temperature range. This conclusion is consistent with the physical picture of the AHE in $CsV_3Sb_5$[12, 16]. Accordingly, we conclude that the third-order nonlinear transport in $CsV_3Sb_5$ arises from a combination of intrinsic quantum geometric contribution and extrinsic skew scattering contribution.

## Conclusion

In summary, we have studied the third-order nonlinear longitudinal and Hall transport in $CsV_3Sb_5$. Both the third-order nonlinear longitudinal and Hall signals can persist up to room temperature. Two characteristic transition temperature points at 39

K and 77 K are observed, where the third-order nonlinear transport undergoes dramatic changes. It is revealed that 77 K corresponds to the CDW transition of $CsV_3Sb_5$, while 39 K is associated with the anomalous symmetry-breaking. Below $T_{CDW}$, the third-order nonlinear longitudinal transport originates from the combined contribution of the quantum metric quadrupole and extrinsic scattering, and the third-order nonlinear Hall response arises from the joint effects of the Berry curvature quadrupole and extrinsic scattering. In contrast, the third-order nonlinear transport is governed by extrinsic contribution above $T_{CDW}$. Our work advances the understanding of the correlations among quantum geometry, crystalline symmetry and nonlinear transport, and deliver a probe for detecting the anomalous symmetry-breaking in $CsV_3Sb_5$.

## Experimental Methods

*Device Fabrication.* The $CsV_3Sb_5$ thin flakes were mechanically exfoliated onto $SiO_2$/Si substrate in a glove box with $H_2O$ and $O_2$ levels below 0.1 ppm. The Au electrodes were fabricated by electron beam lithography and electron beam evaporation. The sample thickness was determined by atomic force microscopy (Cypher S). For protection, hexagonal boron nitride (*h*-BN) thin flake was introduced as capping layer on the $CsV_3Sb_5$ sample.

*Transport Measurements.* Electrical transport measurements were performed in a Quantum Design Physical Property Measurement System (PPMS) with temperatures down to 1.6 K and magnetic field up to 14 T. The voltage difference between different probes was measured by standard lock-in amplifiers (SR 830). The 2-probe DC measurements are performed by using a Keithley 2400. The data shown in the manuscript is collected at a low frequency (47 Hz).

## Acknowledgements

We acknowledge the support of the National Natural Science Foundation of China (Grant Nos. 12374043, 92580203, 12025404, T2221003, 12504053), the National Key R&D Program of China (Grant No. 2022YFA1402404), the Natural Science Foundation of Jiangsu Province (Nos. BK20240166, BK20243013, BK20233001), the Fundamental and Interdisciplinary Disciplines Breakthrough Plan of the Ministry of Education of China (No. JYB2025XDXM411), and the Fundamental Research Funds for the Central Universities (Nos. KG202501, KG202603).

## REFERENCES

(1) Mazin, I. I.; Jeschke, H. O.; Lechermann, F.; Lee, H.; Fink, M.; Thomale, R.; Valentí, R. Theoretical prediction of a strongly correlated Dirac metal. *Nat. Commun.* **2014**, *5* (1), 4261.

(2) Ye, L.; Kang, M.; Liu, J.; von Cube, F.; Wicker, C. R.; Suzuki, T.; Jozwiak, C.; Bostwick, A.; Rotenberg, E.; Bell, D. C.; et al. Massive Dirac fermions in a ferromagnetic kagome metal. *Nature* **2018**, *555* (7698), 638-642.

(3) Kang, M.; Fang, S.; Ye, L.; Po, H. C.; Denlinger, J.; Jozwiak, C.; Bostwick, A.; Rotenberg, E.; Kaxiras, E.; Checkelsky, J. G.; et al. Topological flat bands in frustrated kagome lattice CoSn. *Nat. Commun.* **2020**, *11* (1), 4004.

(4) Kang, M.; Ye, L.; Fang, S.; You, J.-S.; Levitan, A.; Han, M.; Facio, J. I.; Jozwiak, C.; Bostwick, A.; Rotenberg, E.; et al. Dirac fermions and flat bands in the ideal kagome metal FeSn. *Nat. Mater.* **2020**, *19* (2), 163-169.

(5) Liu, Z.; Li, M.; Wang, Q.; Wang, G.; Wen, C.; Jiang, K.; Lu, X.; Yan, S.; Huang, Y.; Shen, D.; et al. Orbital-selective Dirac fermions and extremely flat bands in frustrated kagome-lattice metal CoSn. *Nat. Commun.* **2020**, *11* (1), 4002.

(6) Norman, M. R. Colloquium: Herbertsmithite and the search for the quantum spin liquid. *Rev. Mod. Phys.* **2016**, *88* (4), 041002.

(7) Matthew P. Shores, E. A. N., Bart M. Bartlett, and Daniel G. Nocera. A Structurally Perfect S=1/2 Kagomé Antiferromagnet. *J. Am. Chem. Soc.* **2005**, *127* (39), 13462–13463.

(8) Balents, L. Spin liquids in frustrated magnets. *Nature* **2010**, *464* (7286), 199-208.

(9) Ortiz, B. R.; Gomes, L. C.; Morey, J. R.; Winiarski, M.; Bordelon, M.; Mangum, J. S.; Oswald, I. W. H.; Rodriguez-Rivera, J. A.; Neilson, J. R.; Wilson, S. D.; et al. New kagome prototype materials: discovery of $KV_3Sb_5$, $RbV_3Sb_5$, and $CsV_3Sb_5$. *Phys. Rev. Mater.* **2019**, *3* (9), 094407.

(10) Fu, Y.; Zhao, N.; Chen, Z.; Yin, Q.; Tu, Z.; Gong, C.; Xi, C.; Zhu, X.; Sun, Y.; Liu, K.; et al. Quantum Transport Evidence of Topological Band Structures of Kagome Superconductor $CsV_3Sb_5$. *Phys. Rev. Lett.* **2021**, *127* (20), 207002.

(11) Ortiz, B. R.; Teicher, S. M. L.; Kautzsch, L.; Sarte, P. M.; Ratcliff, N.; Harter, J.; Ruff, J. P. C.; Seshadri, R.; Wilson, S. D. Fermi Surface Mapping and the Nature of Charge-Density-Wave Order in the Kagome Superconductor $CsV_3Sb_5$. *Phys. Rev. X* **2021**, *11* (4), 041030.

(12) Yu, F. H.; Wu, T.; Wang, Z. Y.; Lei, B.; Zhuo, W. Z.; Ying, J. J.; Chen, X. H. Concurrence of anomalous Hall effect and charge density wave in a superconducting topological kagome metal. *Phys. Rev. B* **2021**, *104* (4), L041103.

(13) Ortiz, B. R.; Teicher, S. M. L.; Hu, Y.; Zuo, J. L.; Sarte, P. M.; Schueller, E. C.; Abeykoon, A. M. M.; Krogstad, M. J.; Rosenkranz, S.; Osborn, R.; et al. $CsV_3Sb_5$ : A $Z_2$ Topological Kagome Metal with a Superconducting Ground State. *Phys. Rev. Lett.* **2020**, *125* (24), 247002.

(14) Xu, Z.; Le, T.; Lin, X. $AV_3Sb_5$ Kagome Superconductors: A Review with Transport Measurements. *Chin. Phys. Lett.* **2025**, *42* (3), 037304.

(15) Liu, Y.; Liu, C.-C.; Zhu, Q.-Q.; Ji, L.-W.; Wu, S.-Q.; Sun, Y.-L.; Bao, J.-K.; Jiao, W.-H.; Xu, X.-F.; Ren, Z.; et al. Enhancement of superconductivity and suppression of charge-density wave in As-doped $CsV_3Sb_5$. *Phys. Rev. Mater.* **2022**, *6* (12), 124803.
(16) Zheng, G.; Tan, C.; Chen, Z.; Wang, M.; Zhu, X.; Albarakati, S.; Algarni, M.; Partridge, J.; Farrar, L.; Zhou, J.; et al. Electrically controlled superconductor-to failed insulator transition and giant anomalous Hall effect in kagome metal $CsV_3Sb_5$ nanoflakes. *Nat. Commun.* **2023**, *14* (1), 678.
(17) Yu, F. H.; Ma, D. H.; Zhuo, W. Z.; Liu, S. Q.; Wen, X. K.; Lei, B.; Ying, J. J.; Chen, X. H. Unusual competition of superconductivity and charge-density-wave state in a compressed topological kagome metal. *Nat. Commun.* **2021**, *12* (1), 3645.
(18) Song, Y.; Ying, T.; Chen, X.; Han, X.; Wu, X.; Schnyder, A. P.; Huang, Y.; Guo, J.-g.; Chen, X. Competition of Superconductivity and Charge Density Wave in Selective Oxidized $CsV_3Sb_5$ Thin Flakes. *Phys. Rev. Lett.* **2021**, *127* (23), 237001.
(19) Song, B.; Ying, T.; Wu, X.; Xia, W.; Yin, Q.; Zhang, Q.; Song, Y.; Yang, X.; Guo, J.; Gu, L.; et al. Anomalous enhancement of charge density wave in kagome superconductor $CsV_3Sb_5$ approaching the 2D limit. *Nat. Commun.* **2023**, *14* (1), 2492.
(20) Chen, H.; Yang, H.; Hu, B.; Zhao, Z.; Yuan, J.; Xing, Y.; Qian, G.; Huang, Z.; Li, G.; Ye, Y.; et al. Roton pair density wave in a strong-coupling kagome superconductor. *Nature* **2021**, *599* (7884), 222-228.
(21) Xiang, Y.; Li, Q.; Li, Y.; Xie, W.; Yang, H.; Wang, Z.; Yao, Y.; Wen, H.-H. Twofold symmetry of c-axis resistivity in topological kagome superconductor $CsV_3Sb_5$ with in-plane rotating magnetic field. *Nat. Commun.* **2021**, *12* (1), 6727.
(22) Xu, Y.; Ni, Z.; Liu, Y.; Ortiz, B. R.; Deng, Q.; Wilson, S. D.; Yan, B.; Balents, L.; Wu, L. Three-state nematicity and magneto-optical Kerr effect in the charge density waves in kagome superconductors. *Nat. Phys.* **2022**, *18* (12), 1470-1475.
(23) Wu, Q.; Wang, Z. X.; Liu, Q. M.; Li, R. S.; Xu, S. X.; Yin, Q. W.; Gong, C. S.; Tu, Z. J.; Lei, H. C.; Dong, T.; et al. Simultaneous formation of two-fold rotation symmetry with charge order in the kagome superconductor $CsV_3Sb_5$ by optical polarization rotation measurement. *Phys. Rev. B* **2022**, *106* (20), 205109.
(24) Yu, F.-H.; Wen, X.-K.; Gui, Z.-G.; Wu, T.; Wang, Z.; Xiang, Z.-J.; Ying, J.; Chen, X. Pressure tuning of the anomalous Hall effect in the kagome superconductor $CsV_3Sb_5$. *Chinese Physics B* **2022**, *31* (1), 017405.
(25) Wang, L.; Zhang, W.; Wang, Z.; Poon, T. F.; Wang, W.; Tsang, C. W.; Xie, J.; Zhou, X.; Zhao, Y.; Wang, S.; et al. Anomalous Hall effect and two-dimensional Fermi surfaces in the charge-density-wave state of kagome metal $RbV_3Sb_5$. *Journal of Physics: Materials* **2023**, *6* (2), 02lt01.
(26) Liu, S.; Roppongi, M.; Kimata, M.; Ishihara, K.; Grasset, R.; Konczykowski, M.; Ortiz, B. R.; Wilson, S. D.; Yoshimi, K.; Shibauchi, T.; et al. Impact of Tiny Fermi Pockets with Extremely High Mobility on the Hall Anomaly in the Kagome Metal $CsV_3Sb_5$. *Phys. Rev. Lett.* **2025**, *135* (5), 056502.
(27) Deng, H.; Qin, H.; Liu, G.; Yang, T.; Fu, R.; Zhang, Z.; Wu, X.; Wang, Z.; Shi, Y.; Liu, J.; et al. Chiral kagome superconductivity modulations with residual Fermi arcs. *Nature* **2024**, *632* (8026), 775-781.

(28) Ge, J.; Wang, P.; Xing, Y.; Yin, Q.; Wang, A.; Shen, J.; Lei, H.; Wang, Z.; Wang, J. Charge-4e and Charge-6e Flux Quantization and Higher Charge Superconductivity in Kagome Superconductor Ring Devices. *Phys. Rev. X* **2024**, *14* (2), 021025.
(29) Wu, Y.; Wang, Q.; Zhou, X.; Wang, J.; Dong, P.; He, J.; Ding, Y.; Teng, B.; Zhang, Y.; Li, Y.; et al. Nonreciprocal charge transport in topological kagome superconductor $CsV_3Sb_5$. *npj Quantum Mater.* **2022**, *7* (1), 105.
(30) Le, T.; Pan, Z.; Xu, Z.; Liu, J.; Wang, J.; Lou, Z.; Yang, X.; Wang, Z.; Yao, Y.; Wu, C.; et al. Superconducting diode effect and interference patterns in kagome $CsV_3Sb_5$. *Nature* **2024**, *630* (8015), 64-69.
(31) Ge, J.; Liu, X.; Wang, P.; Pang, H.; Yin, Q.; Lei, H.; Wang, Z.; Wang, J. Nonreciprocal superconducting critical currents with normal state field trainability in kagome superconductor $CsV_3Sb_5$. *Nat. Commun.* **2026**, *17* (1), 6056.
(32) Kang, M.; Fang, S.; Kim, J.-K.; Ortiz, B. R.; Ryu, S. H.; Kim, J.; Yoo, J.; Sangiovanni, G.; Di Sante, D.; Park, B.-G.; et al. Twofold van Hove singularity and origin of charge order in topological kagome superconductor $CsV_3Sb_5$. *Nat. Phys.* **2022**, *18* (3), 301-308.
(33) Luo, J.; Zhao, Z.; Zhou, Y. Z.; Yang, J.; Fang, A. F.; Yang, H. T.; Gao, H. J.; Zhou, R.; Zheng, G.-q. Possible star-of-David pattern charge density wave with additional modulation in the kagome superconductor $CsV_3Sb_5$. *npj Quantum Mater.* **2022**, *7* (1), 30.
(34) Mielke, C.; Das, D.; Yin, J. X.; Liu, H.; Gupta, R.; Jiang, Y. X.; Medarde, M.; Wu, X.; Lei, H. C.; Chang, J.; et al. Time-reversal symmetry-breaking charge order in a kagome superconductor. *Nature* **2022**, *602* (7896), 245-250.
(35) Gui, H.; Yang, L.; Wang, X.; Chen, D.; Shi, Z.; Zhang, J.; Wei, J.; Zhou, K.; Schnelle, W.; Zhang, Y.; et al. Probing orbital magnetism of a kagome metal $CsV_3Sb_5$ by a tuning fork resonator. *Nat. Commun.* **2025**, *16* (1), 4275.
(36) Hao, J.; Zhou, X.; Li, Y.; Liu, Z.; Ji, B.; Dai, Y.; Wang, Z.; Wen, H.-H. Low-energy gap associated with the anomalous symmetry-breaking states in $CsV_3Sb_5$. *Phys. Rev. B* **2026**, *113* (3), 035128.
(37) Xu, L.; Xie, Z.; Wang, J.; Yin, Q.; Lei, H.; Zhang, J. Geometry-Enhanced Second-Harmonic Charge Transport in Kagome Superconductor $CsV_3Sb_5$. *Nano Lett.* **2025**, *25* (38), 14222-14228.
(38) Nie, L.; Sun, K.; Ma, W.; Song, D.; Zheng, L.; Liang, Z.; Wu, P.; Yu, F.; Li, J.; Shan, M.; et al. Charge-density-wave-driven electronic nematicity in a kagome superconductor. *Nature* **2022**, *604* (7904), 59-64.
(39) Chen, D.; He, B.; Yao, M.; Pan, Y.; Lin, H.; Schnelle, W.; Sun, Y.; Gooth, J.; Taillefer, L.; Felser, C. Anomalous thermoelectric effects and quantum oscillations in the kagome metal $CsV_3Sb_5$. *Phys. Rev. B* **2022**, *105* (20), L201109.
(40) Zhao, H.; Li, H.; Ortiz, B. R.; Teicher, S. M. L.; Park, T.; Ye, M.; Wang, Z.; Balents, L.; Wilson, S. D.; Zeljkovic, I. Cascade of correlated electron states in the kagome superconductor $CsV_3Sb_5$. *Nature* **2021**, *599* (7884), 216-221.
(41) Suetsugu, S.; Hori, F.; Shibata, M.; Kitagawa, S.; Ishida, K.; Asaba, T.; Nakazawa, S.; Li, Q.; Wen, H. H.; Shibauchi, T.; et al. Microscopic signatures of an imaginary

charge density wave in a kagome metal. *Nat. Phys.* **2026**, doi : 10.1038/s41567-41026-03339-41568.
(42) Cha, J.; Lee, H.; Sim, S.; Sur, Y.; Kim, K.-T.; Han, J.-H.; Kim, S.-W.; Lee, G.; Hyun, J.; Lim, C.-y.; et al. Evidence of time-reversal symmetry breaking above the charge density wave order in a kagome metal. *Nat. Phys.* **2026**, doi : 10.1038/s41567-41026-03331-41562.
(43) Frachet, M.; Wang, L.; Xia, W.; Guo, Y.; He, M.; Maraytta, N.; Heid, R.; Haghighirad, A.-A.; Merz, M.; Meingast, C.; et al. Colossal c-Axis Response and Lack of Rotational Symmetry Breaking within the Kagome Planes of the $CsV_3Sb_5$ Superconductor. *Phys. Rev. Lett.* **2024**, *132* (18), 186001.
(44) Feng, X. Y.; Zhao, Z.; Luo, J.; Zhou, Y. Z.; Yang, J.; Fang, A. F.; Yang, H. T.; Gao, H. J.; Zhou, R.; Zheng, G.-q. Fully-gapped superconductivity with rotational symmetry breaking in pressurized kagome metal $CsV_3Sb_5$. *Nat. Commun.* **2025**, *16* (1), 3643.
(45) Guo, C.; Putzke, C.; Konyzheva, S.; Huang, X.; Gutierrez-Amigo, M.; Errea, I.; Chen, D.; Vergniory, M. G.; Felser, C.; Fischer, M. H.; et al. Switchable chiral transport in charge-ordered kagome metal $CsV_3Sb_5$. *Nature* **2022**, *611* (7936), 461-466.
(46) Guo, C.; Wagner, G.; Putzke, C.; Chen, D.; Wang, K.; Zhang, L.; Gutierrez-Amigo, M.; Errea, I.; Vergniory, M. G.; Felser, C.; et al. Correlated order at the tipping point in the kagome metal $CsV_3Sb_5$. *Nat. Phys.* **2024**, *20* (4), 579-584.
(47) Wei, X.; Tian, C.; Cui, H.; Zhai, Y.; Li, Y.; Liu, S.; Song, Y.; Feng, Y.; Huang, M.; Wang, Z.; et al. Three-dimensional hidden phase probed by in-plane magnetotransport in kagome metal $CsV_3Sb_5$ thin flakes. *Nat. Commun.* **2024**, *15* (1), 5038.
(48) Guo, C.; Wang, K.; Zhang, L.; Putzke, C.; Chen, D.; van Delft, M. R.; Wiedmann, S.; Balakirev, F. F.; McDonald, R. D.; Gutierrez-Amigo, M.; et al. Many-body interference in kagome crystals. *Nature* **2025**, *647* (8088), 68-73.
(49) Gao, Y.; Yang, S. A.; Niu, Q. Field Induced Positional Shift of Bloch Electrons and Its Dynamical Implications. *Phys. Rev. Lett.* **2014**, *112* (16), 166601.
(50) Sodemann, I.; Fu, L. Quantum Nonlinear Hall Effect Induced by Berry Curvature Dipole in Time-Reversal Invariant Materials. *Phys. Rev. Lett.* **2015**, *115* (21), 216806.
(51) Wang, N.; You, J.-Y.; Wang, A.; Zhou, X.; Zhang, Z.; Lai, S.; Feng, Y.-P.; Lin, H.; Chang, G.; Gao, W.-b. Non-centrosymmetric topological phase probed by non-linear Hall effect. *Natl. Sci. Rev.* **2024**, *11* (6), nwad103.
(52) Wang, E.; Zeng, H.; Duan, W.; Huang, H. Spontaneous Inversion Symmetry Breaking and Emergence of Berry Curvature and Orbital Magnetization in Topological $ZrTe_5$ Films. *Phys. Rev. Lett.* **2024**, *132* (26), 266802.
(53) Wang, Y.; Legg, H. F.; Bomerich, T.; Park, J.; Biesenkamp, S.; Taskin, A. A.; Braden, M.; Rosch, A.; Ando, Y. Gigantic Magnetochiral Anisotropy in the Topological Semimetal $ZrTe_5$. *Phys. Rev. Lett.* **2022**, *128* (17), 176602.
(54) Dixit, A.; Sivakumar, P. K.; Manna, K.; Felser, C.; Parkin, S. S. P. A chiral fermionic valve driven by quantum geometry. *Nature* **2025**, *649* (8095), 47-52.
(55) Jiang, H.; Xi, T.; Li, J.; He, Y.; Ma, H.; Mao, Y.; Taniguchi, T.; Watanabe, K.; Rhodes, D. A.; Zhang, Y.; et al. Probing interplay of topological properties and electron correlation in $TaIrTe_4$ via nonlinear Hall effect. *Nat. Commun.* **2025**, *16* (1), 6351.

(56) Zhao, D.; Li, Z.; Sun, J.; Zhao, Y.; Wang, T.; Qi, L.; Tang, W.; Zhang, S.; Ramiere, A.; Jin, H.; et al. Charge Density Wave-Induced Highly Sensitive Terahertz Detection Based on a Large Nonlinear Hall Effect. *ACS Nano* **2026**, *20* (22), 16325-16335.
(57) Chen, Z.-H.; Liao, X.; Dong, J.-W.; Liu, X.-Y.; Zhao, T.-Y.; Li, D.; Wang, A.-Q.; Liao, Z.-M. Charge density wave modulated third-order nonlinear Hall effect in 1T−$VSe_2$ nanosheets. *Phys. Rev. B* **2024**, *110* (23), 235135.
(58) Sankar, S.; Liu, R.; Zhang, C.-P.; Li, Q.-F.; Chen, C.; Gao, X.-J.; Zheng, J.; Lin, Y.-H.; Qian, K.; Yu, R.-P.; et al. Experimental Evidence for a Berry Curvature Quadrupole in an Antiferromagnet. *Phys. Rev. X* **2024**, *14* (2), 021046.
(59) Li, H.; Zhang, C.; Zhou, C.; Ma, C.; Lei, X.; Jin, Z.; He, H.; Li, B.; Law, K. T.; Wang, J. Quantum geometry quadrupole-induced third-order nonlinear transport in antiferromagnetic topological insulator $MnBi_2Te_4$. *Nat. Commun.* **2024**, *15* (1), 7779.
(60) Li, X.; Dai, Z.; Zhang, S.; Zhang, H.; Li, C.; Wei, B.; Guo, F.; Li, C.; Fei, F.; Zhang, M.; et al. Higher odd-order nonlinear Hall effect in magnetic topological insulator $Mn(Bi_{1-x}Sb_x)_2Te_4$. *Nat. Commun.* **2026**, *17* (1), 6652.
(61) Li, S.; Wang, X.; Yang, Z.; Zhang, L.; Teo, S. L.; Lin, M.; He, R.; Wang, N.; Song, P.; Tian, W.; et al. Giant Third-Order Nonlinear Hall Effect in Misfit Layer Compound $(SnS)_{1.17}(NbS_2)_3$. *ACS Appl. Mater. Interfaces* **2024**, *16* (8), 11043-11049.
(62) Khasanov, R.; Das, D.; Gupta, R.; Mielke, C.; Elender, M.; Yin, Q.; Tu, Z.; Gong, C.; Lei, H.; Ritz, E. T.; et al. Time-reversal symmetry broken by charge order in $CsV_3Sb_5$. *Phys. Rev. Res.* **2022**, *4* (2), 023244.
(63) Wang, C.; Xiao, R. C.; Liu, H.; Zhang, Z.; Lai, S.; Zhu, C.; Cai, H.; Wang, N.; Chen, S.; Deng, Y.; et al. Room-temperature third-order nonlinear Hall effect in Weyl semimetal $TaIrTe_4$. *Natl. Sci. Rev.* **2022**, *9* (12), nwac020.
(64) Guo, Y.; Jiang, C.; Song, J.; Zhang, K.; Qiu, D.; Yang, C.; Wang, Y.; Wang, H.; Li, Y.; Zhang, X.; et al. Decoupling Surface and Bulk States via Third Order Electrical Nonlinearity in Centrosymmetric Crystal. *Nano Lett.* **2025**, *25* (47), 16621–16629.
(65) Tian, Y.; Ye, L.; Jin, X. Proper Scaling of the Anomalous Hall Effect. *Phys. Rev. Lett.* **2009**, *103* (8), 087206.
(66) Kang, K.; Li, T.; Sohn, E.; Shan, J.; Mak, K. F. Nonlinear anomalous Hall effect in few-layer $WTe_2$. *Nat. Mater.* **2019**, *18* (4), 324-328.
(67) Liu, X.-Y.; Wang, A.-Q.; Li, D.; Zhao, T.-Y.; Liao, X.; Liao, Z.-M. Giant Third-Order Nonlinearity Induced by the Quantum Metric Quadrupole in Few-Layer $WTe_2$. *Phys. Rev. Lett.* **2025**, *134* (2), 026305.
(68) Chen, H.; Qin, P.; Meng, Z.; Zhao, G.; Chen, K.; Xi, C.; Wang, X.; Liu, L.; Duan, Z.; Jiang, S.; et al. Giant room-temperature third-order electrical transport in a thin-film altermagnet candidate. *Nat. Nanotechnol.* **2026**, *21* (6), 779-785.

## Figures and Captions

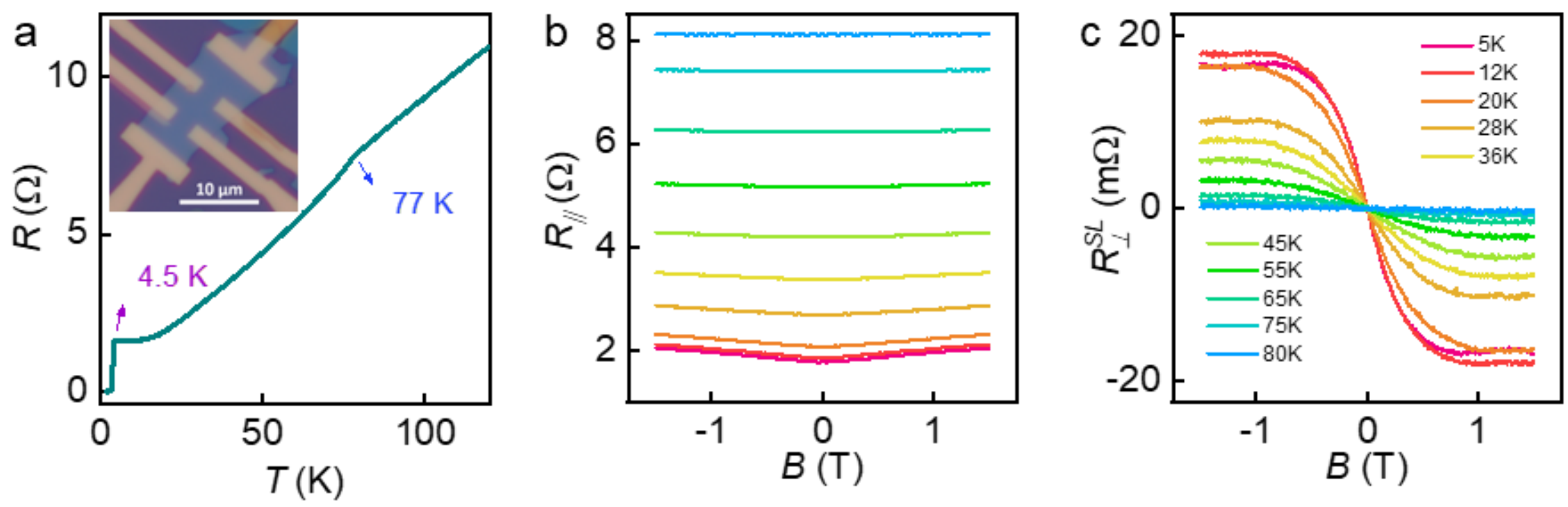

FIG. 1. Basic characterization of the $CsV_3Sb_5$ sample. (a) The temperature-dependent longitudinal resistance at zero magnetic fields. Inset is the optical image of the $CsV_3Sb_5$ sample. The white scale bar is 10 μm. The thickness of $CsV_3Sb_5$ sample is 22.7 nm (b)-(c) The magnetic-field-dependent longitudinal resistance $R_{\parallel}$ and step-like Hall resistance $R_{\perp}^{SL}$ at temperatures ranging from 5 K to 80 K.

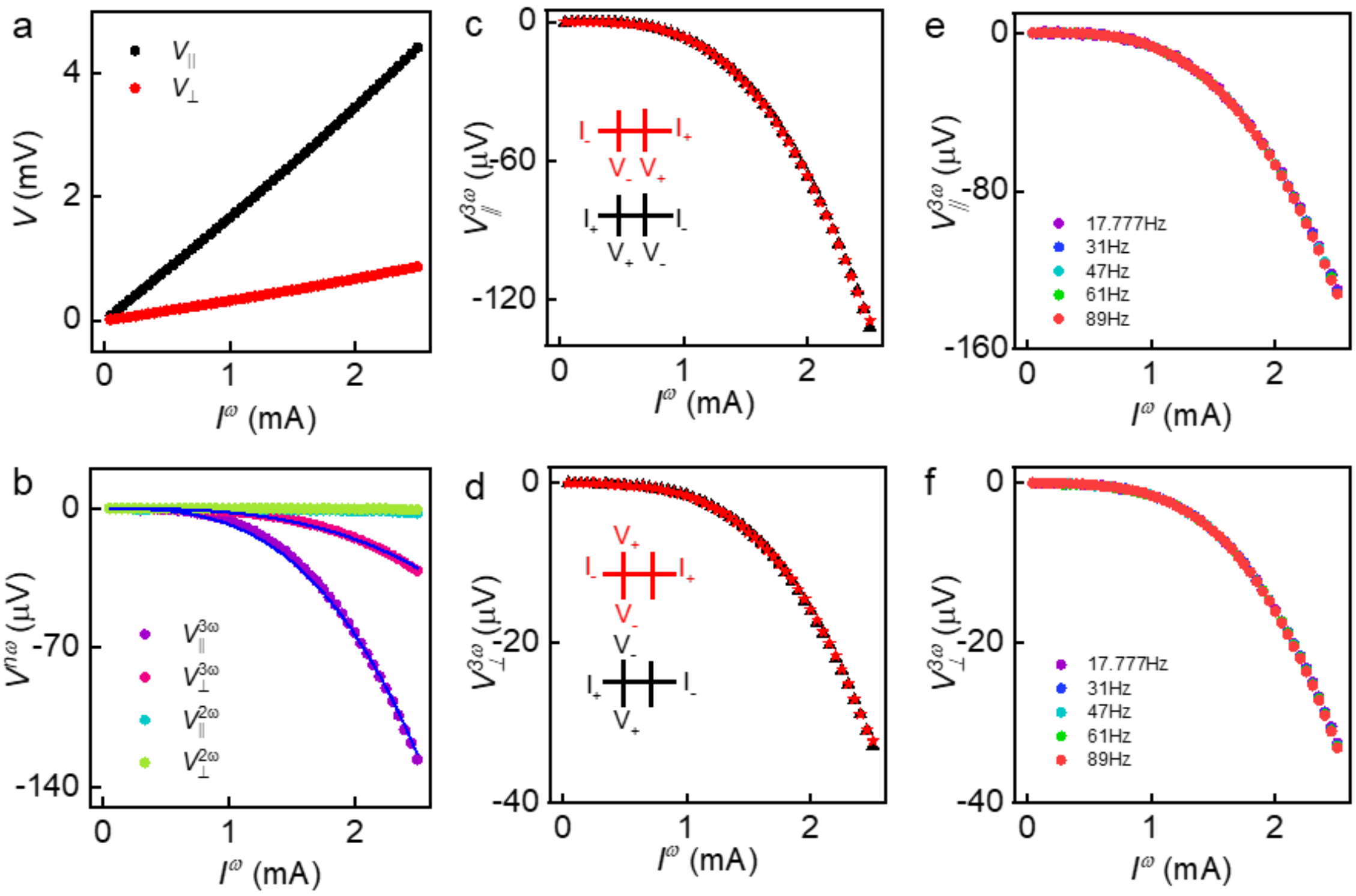

FIG. 2. The third-order nonlinear transport of $CsV_3Sb_5$ at 5 K. (a) The first-order longitudinal and Hall voltages as a function of the current $I^{\omega}$. (b) The second- and third-order nonlinear voltages in the longitudinal and Hall directions as a function of the $I^{\omega}$. The blue solid lines are cubic fits. (c)-(d) The third-order nonlinear longitudinal and Hall voltages as a function of the $I^{\omega}$ under different configuration. The insets illustrate the measurement configurations, in which both the current direction and the voltage-lead connections are reversed. (e)-(f) The third-order nonlinear longitudinal and Hall voltages under different frequencies.

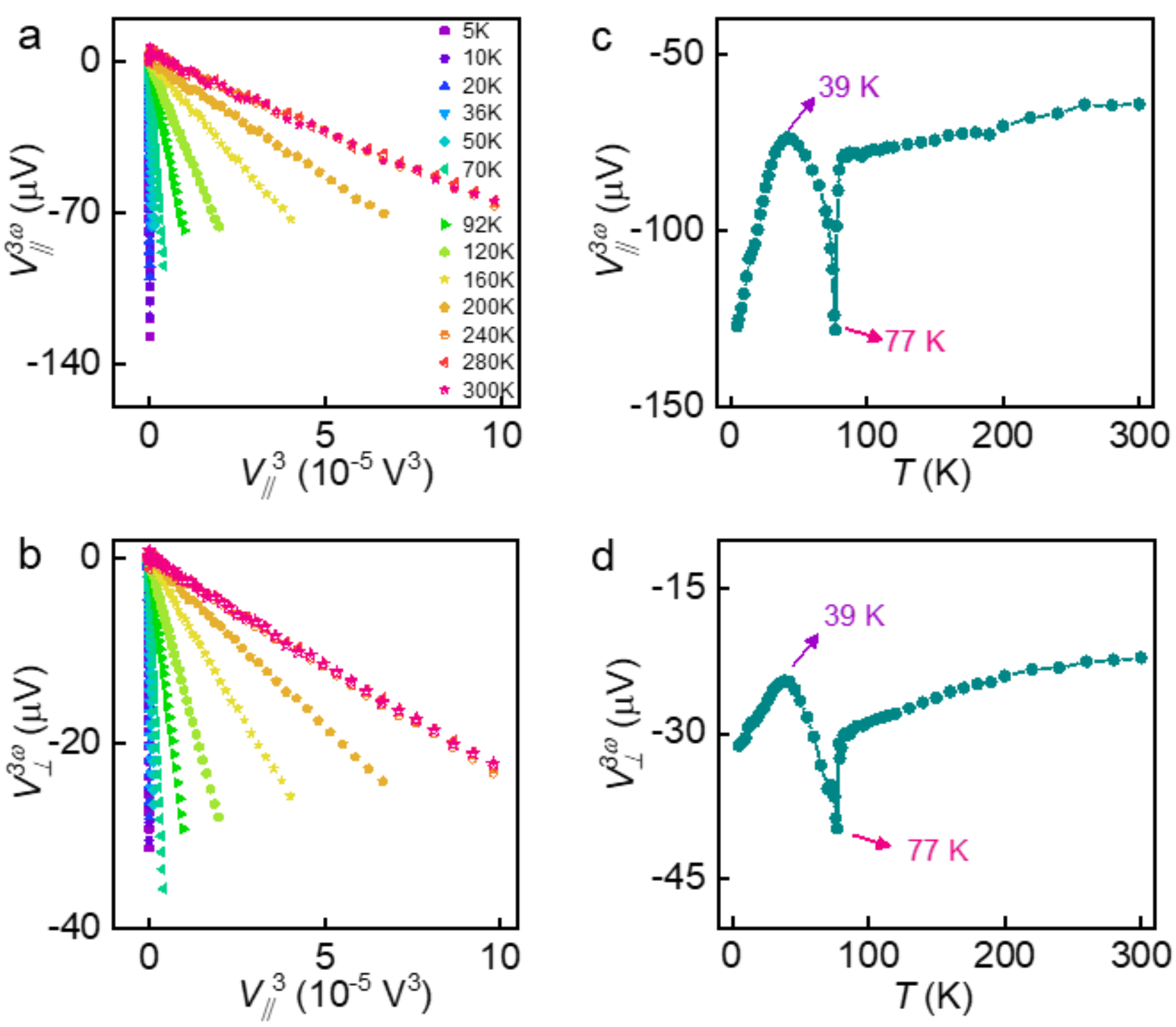

FIG. 3. The temperature-dependent third-order nonlinearity of $CsV_3Sb_5$. (a)-(b) The third-order nonlinear longitudinal voltage $V_{\parallel}^{3\omega}$ and Hall voltage $V_{\perp}^{3\omega}$ as a function of the cube of the longitudinal voltage $V_{\parallel}$ under different temperatures from 5 K to 300 K at $B = 0$ T. (c)-(d) The temperature evolution of the third-order nonlinear longitudinal and Hall voltages with the current of 2.5 mA.

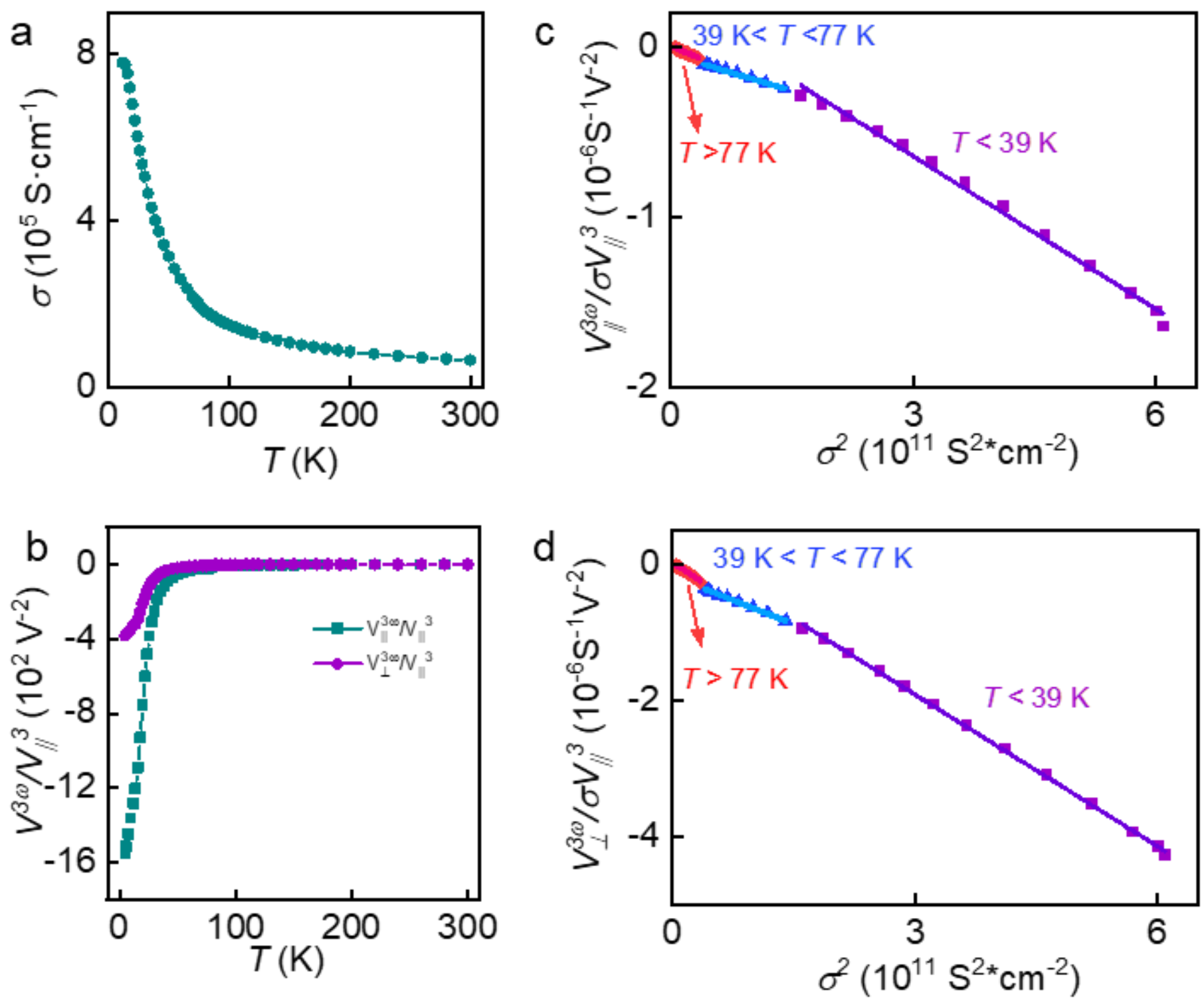

FIG. 4. The scaling law of the third-order nonlinearity in $CsV_3Sb_5$. (a) The temperature-dependent longitudinal conductivity $\sigma$. (b) The ratio of the third-order nonlinear longitudinal and Hall voltages to the cubed longitudinal voltages as a function of temperature. (c)-(d) The $V_{\parallel}^{3\omega}/\sigma V_{\parallel}^{3}$, $V_{\perp}^{3\omega}/\sigma V_{\parallel}^{3}$ and fit to the data plotted with the square of $\sigma$. The scatter symbols denote experimental data and the solid lines represent the fitting results.